# Alternative metrics in scientometrics:

# A meta-analysis of research into three altmetrics

Lutz Bornmann

Division for Science and Innovation Studies

Administrative Headquarters of the Max Planck Society

Hofgartenstr. 8,

80539 Munich, Germany.

Email: bornmann@gv.mpg.de

**Abstract**

Alternative metrics are currently one of the most popular research topics in scientometric research. This paper provides an overview of research into three of the most important altmetrics: microblogging (Twitter), online reference managers (Mendeley and CiteULike) and blogging. The literature is discussed in relation to the possible use of altmetrics in research evaluation. Since the research was particularly interested in the correlation between altmetrics counts and citation counts, this overview focuses particularly on this correlation. For each altmetric, a meta-analysis is calculated for its correlation with traditional citation counts. As the results of the meta-analyses show, the correlation with traditional citations for micro-blogging counts is negligible (pooled r=0.003), for blog counts it is small (pooled r=0.12) and for bookmark counts from online reference managers, medium to large (CiteULike pooled r=0.23; Mendeley pooled r=0.51).

# 1 Introduction

The interdisciplinary Science Citation Index, developed in the 1950s and published on a regular basis since the 1960s, was firstly considered as an evaluation metric when it was included in the US National Science Foundation's Science Indicators Reports in 1972 (de Bellis, 2014). It was not until a few years later that research evaluation got to grips with this development, when researchers approached the new bibliometric data from the point of view of theory and empirically-statistically. For example, in 1978 the journal *Scientometrics* was launched, and the Information Science and Scientometric Research Unit (ISSRU) at the Hungarian Academy of Sciences established in Budapest. Scientometrics established itself as an independent discipline, with its own conferences, its own professorships and developed its own instruments for the assessment of the data. The scientometric research on citation data was particularly concerned with the theoretical basis of citation processes (Wouters, 2014), the citation behaviour of researchers (Bornmann & Daniel, 2008), the development of advanced indicators (in particular normalised indicators which can be used independently of time and subject area) (Bornmann, Leydesdorff, & Mutz, 2013) and the visualisation of bibliometric data in science maps or networks (Börner, Sanyal, & Vespignani, 2007; Bornmann, Stefaner, de Moya Anegón, & Mutz, 2014). It was not until the development of normalised indicators, which began in the mid-1980s (Schubert & Braun, 1986), first allowed the use of citation data for the formulation of meaningful results in research evaluation.

In recent years, scientometrics seems to be situated in a similar point of development as in the 1960s. Social media platforms, such as for example Mendeley or Twitter, make data available (known as altmetrics), which allow the impact of research to be measured more broadly than with citations (Priem, 2014). This offer appeared almost simultaneously with the wish in science policy to be informed not only about the impact of research on research itself, but also of research on other segments of society (Bornmann, 2012, 2013). So it can be

expected – as in 1972 with the incorporation of citation data in the Science Indicators Report – that altmetrics will very soon be incorporated in research evaluation (Colledge, 2014). The use of altmetrics in research evaluation will probably happen before the point in time when – similarly to the mid-1980s – advanced altmetric indicators are suggested with the necessary normalisation. Altmetric.com – a start-up company which offers altmetrics to publications – recently presented a tool (Altmetric for Institutions), which can represent the broad impact of the research of institutions.

This paper provides an overview of research into three of the most important altmetrics: microblogging (Twitter), online reference managers (Mendeley and CiteULike) and blogging. In recent year, many other types of data have been used in altmetrics, i.e. data from social media like Facebook or Google+; accessing content data from repositories like GitHub, Figshare, Slideshare, Vimeo, or YouTube; domain specific data from arXiv or Dyrad; access measures on publishers sites from PLoS, and amount of comments on a paper from some publishers, like PLoS or BioMed Central. However, this study focuses on microblogging, online reference managers, and blogging, because only these have sufficient amount of published data for meta-analysis. In this study for each of these metrics, a meta-analysis (Glass, 1976) was performed on its correlation with traditional citation counts. For this, the corresponding correlation coefficients were taken from a range of studies. The meta-analysis allows a generalised statement on the correlation between an alternative metric and citations.

In the following, the literature is discussed in relation to the possible use of altmetrics in research assessment. Since the research was particularly interested in the correlation between altmetric counts and citation counts, this overview focuses particularly on this correlation. An altmetric measure with high correlation could be seen as “not random” (Thelwall, 2014), but as a ‘not so’ alternative metric. A low correlation could mean that these measures involve other dimensions of impact than traditional metrics (Dinsmore, Allen, &

Dolby, 2014). Studies of correlation appear to be frequently done because they are easily produced, not because the correlation between citation counts and altmetrics is the most pertinent question to examine. „Altmetrics are not intended to replace traditional bibliometrics like number of citations – in fact, these two approaches are complementary and capture different types of impact for different audiences" (Darling, Shiffman, Côté, & Drew, 2013). However, while correlation between altmetrics and traditional citations is a common research topic, research is also interested in determining the scholarly value of altmetrics and citations (e.g. how they are different and what they measure that is useful or not). Thus, an overview of these other studies is also given in the following sections besides the meta-analysis.

# 2 Methods

## 2.1 Literature search

The literature research was conducted at the beginning of 2014. In a first step, some studies were located that dealt with altmetrics, using the reference lists provided by narrative reviews of research into altmetrics (Bar-Ilan, Shema, & Thelwall, 2014; Galloway, Pease, & Rauh, 2013; Haustein, 2014; Priem, 2014; Rodgers & Barbrow, 2013; Torres-Salinas, Cabezas-Clavijo, & Jimenez-Contreras, 2013) and using tables of contents of certain journals (e.g. *Scientometrics*, *Journal of Informetrics*, and *Journal of the Association for Information Science and Technology*). A search was conducted of both publications (journal articles, monographs, etc.) and grey literature (Internet documents, institutional reports, case reports, etc.) to avoid a bias that follows from the difficulties of publishing non-conforming studies (Eagly, 2005).

In a second step, in order to obtain keywords for searching literature databases, a bibliogram was prepared (White, 2005) for the studies located in the first step. The bibliogram ranks by frequency the words included in the abstracts of the studies located. Words at the top of the ranking list (e.g., altmetrics, Twitter) were used for searches in computerised literature databases (e.g. Web of Science, PubMed, and Scopus) and Internet search engines (e.g. Google Scholar).

In the final step of the literature search, all of the citing publications were located for a series of papers found in the first and second step for which there are a fairly large number of citations in Web of Science.

## 2.2 Meta-analysis

Overviews on particular topics normally use narrative techniques without attempting quantitative synthesis of study results. As, from the viewpoint of quantitative social scientists, narrative reviews are not very precise in their descriptions of study results (Shadish, Cook, & Campbell, 2002), quantitative techniques should be used as well as narrative techniques. The term “meta-analysis” (Glass, 1976) refers to a statistical approach that combines evidence from different studies to obtain an overall estimate of treatment effects. Taking the quantitative approach, meta-analysis allows generalised statements on the strength of the effects, regardless of the specificity of individual studies (Matt & Navarro, 1997). Undertaking a meta-analysis presupposes that each study is similarly designed with regard to certain properties (e.g., methods or sampling). Even though the studies that investigate the relationship between traditional citations and altmetrics are quite heterogeneous, most of them are similar in reporting the correlation coefficient. Using these coefficients, it is possible to evaluate the empirical studies meta-analytically.

In this study, the meta-analyses have been conducted with Stata (StataCorp., 2013) using the metan command (Bradburn, Deeks, & Altman, 1998). In scientometrics, the meta-analysis technique was used, for example, by Wainer and Vieira (2013) to pool the correlations between peer evaluations and different metrics.

# 3 Results

## 3.1 Microblogging (Twitter)

With micro-blogging, users send short messages to other users of a platform. The best known microblogging platform is Twitter (www.twitter.com), which was founded in 2006. Twitter allows the sending of short messages (called tweets) of up to 140 characters (Shema, Bar-Ilan, & Thelwall, 2014). In 2013, Twitter had more than 200 million active users, who sent over 400 million tweets per day (blog.twitter.com/2013/03/celebrating-twitter7.html). About a fifth of all Internet users are active on Twitter (Duggan & Smith, 2014). With a Twitter account one can "follow" other users (and receive their tweets) and can send tweets to one's own followers. Tweets can be categorised with hashtags (#), and other people can be named in tweets with ats (@). A tweet from another person can be forwarded (retweeted) to one's own followers (Darling, et al., 2013; Zubiaga, Spina, Martínez, & Fresno, 2014). Twitter is used mainly for daily chatter, conversations, sharing information and reporting news (Weller, Dröge, & Puschmann, 2011).

While this platform was initially used to communicate in a simple way what one happened to be doing (Priem & Costello, 2010), today it is also often used professionally or for scientific purposes. It is estimated that around one in 40 scientists is active on Twitter (Priem, Costello, & Dzuba, 2012), while marked differences are suspected between disciplines in the number of active users and their usage behaviour (Hammarfelt, 2014; Holmberg & Thelwall, 2014). Whereas the scientists who are registered with Twitter seem to be newer to academia, those who also actively tweet are mainly experienced scientists (Darling, et al., 2013).

With respect to the science-related communication over Twitter, the following terminology is suggested: a tweet which contains or refers to scientific content is referred to as a scientific tweet (Weller, et al., 2011). A subset of the scientific tweets is represented by the twitter citations in which a web link to a peer-reviewed scientific publication is included in the tweet (Priem & Costello, 2010). If a tweet refers directly to a publication, it is described as a first-order citation; if an intermediate web page exists, it is a second-order citation (Priem & Costello, 2010; Weller & Puschmann, 2011). Whereas retweets are described as internal citations, tweets with links to outside information are external citations (Haustein, Peters, Sugimoto, Thelwall, & Larivière, 2014; Weller & Peters, 2012).

Twitter is used by scientists and those interested in science mainly to publicise or to discuss scientific results (and other products of scientific work, such as data sets) and to follow or to comment on live events in science, as conference talks or workshop discussions (Bik & Goldstein, 2013; Holmberg & Thelwall, 2014). Priem and Costello (2010) also suggest that Twitter is used by scientists to obtain pointers to interesting publications (and other products of research). If one follows the right people or institutions, one can very quickly obtain indications via Twitter about recently published research results (Priem, 2014; Puschmann, 2014). Around 40% of Twitter citations already appear within a week of publication (Priem & Costello, 2010; Shuai, Pepe, & Bollen, 2012). According to Darling, et al. (2013) the tweets written about the results of a paper indicate more about the most interesting discoveries or conclusions than the title or the abstract of a paper can.

Twitter citations are seen as an alternative to the traditional metrics for measuring the impact of research. The advantage of Twitter citations consists mainly in their allowing a measurement of the impact of a publication very soon after its appearance (with citations, one must wait at least three years) and that they can provide information about the broader impact. Twitter is used mainly by people outside research. But the use of Twitter citations for research assessment is also subject to criticism.

– The 140-character limit does not allow a user to get to grips with the content of a paper in a tweet (Darling, et al., 2013; Neylon, Willmers, & King, 2014). In addition, Twitter users orientate themselves more to the taste of their followers than to the general usefulness of research (Priem & Costello, 2010). Against this background it is not clear what one is actually measuring with a twitter citation (Gunn, 2013): Is one measuring intellectual impact?

- With Twitter citations it is not clear which audience a research product had. Whereas with traditional citations one knows that scientists have made a reference to a publication, the citing person group is not precisely known for Twitter (Priem, 2014; Weller, et al., 2011).
- Even when a research result is mentioned in a tweet, the link to the corresponding publication is often absent. This leads to underestimation of impact (Mahrt, Weller, & Peters, 2012; Weller, et al., 2011).
- Since only a few researchers use micro-blogging, and many researchers tend to have a negative attitude to micro-blogging as a professional communication platform (Mahrt, et al., 2012), it seems that impact measurement of research on research itself is not possible with Twitter citations. Most of the studies concerned with the use of Twitter by researchers have concluded that "Twitter is not considered a particularly important tool for scholarly dissemination" (Haustein, Peters, Sugimoto, et al., 2014, p. 657).
- Tweets are often copied completely or in part and sent as retweets (Holmberg & Thelwall, 2014). It is not clear how partially copied retweets can be recognised reliably and how the retweets should be handled in alternative impact measurement (Weller & Peters, 2012).

In recent years only a few empirical studies have been carried out on Twitter as a possible source for alternative metrics (Haustein et al., 2014). Haustein, Peters, Sugimoto, et al. (2014) have investigated the share of papers from the Web of Science which have been tweeted at least once. They arrived at the conclusion that this share was around 10% – though with a rising trend in recent years. Similarly to traditional citations, Twitter citations are irregularly distributed across papers, with most papers receiving none or just one citation, and only a few papers receiving a large number. According to the results of Haustein, Peters, Sugimoto, et al. (2014) there are great differences between disciplines in Twitter citation counts: whereas papers from biomedical research have relatively high Twitter citation counts, the counts in physics and mathematics tend to be low.

Darling, et al. (2013) have produced a content analysis of the Twitter profiles of marine scientists who actively tweet about ocean science and conservation. As the results show, their followers consist mainly of scientists and scholarly organisations, as well as non-scientists, non-governmental organisations and media representatives: “The majority of our followers (~55%) comprised science students, scientists or scientific organisations that could be potential collaborators for most scientists. The remaining 45% comprised non-scientists, media and the general public who may be more likely to be engaged in the dissemination of published scientific findings” (Darling, et al., 2013). The marine scientists could not only reach scientists with their research, but also made an impact on non-governmental organisations, private industry, and government agencies.

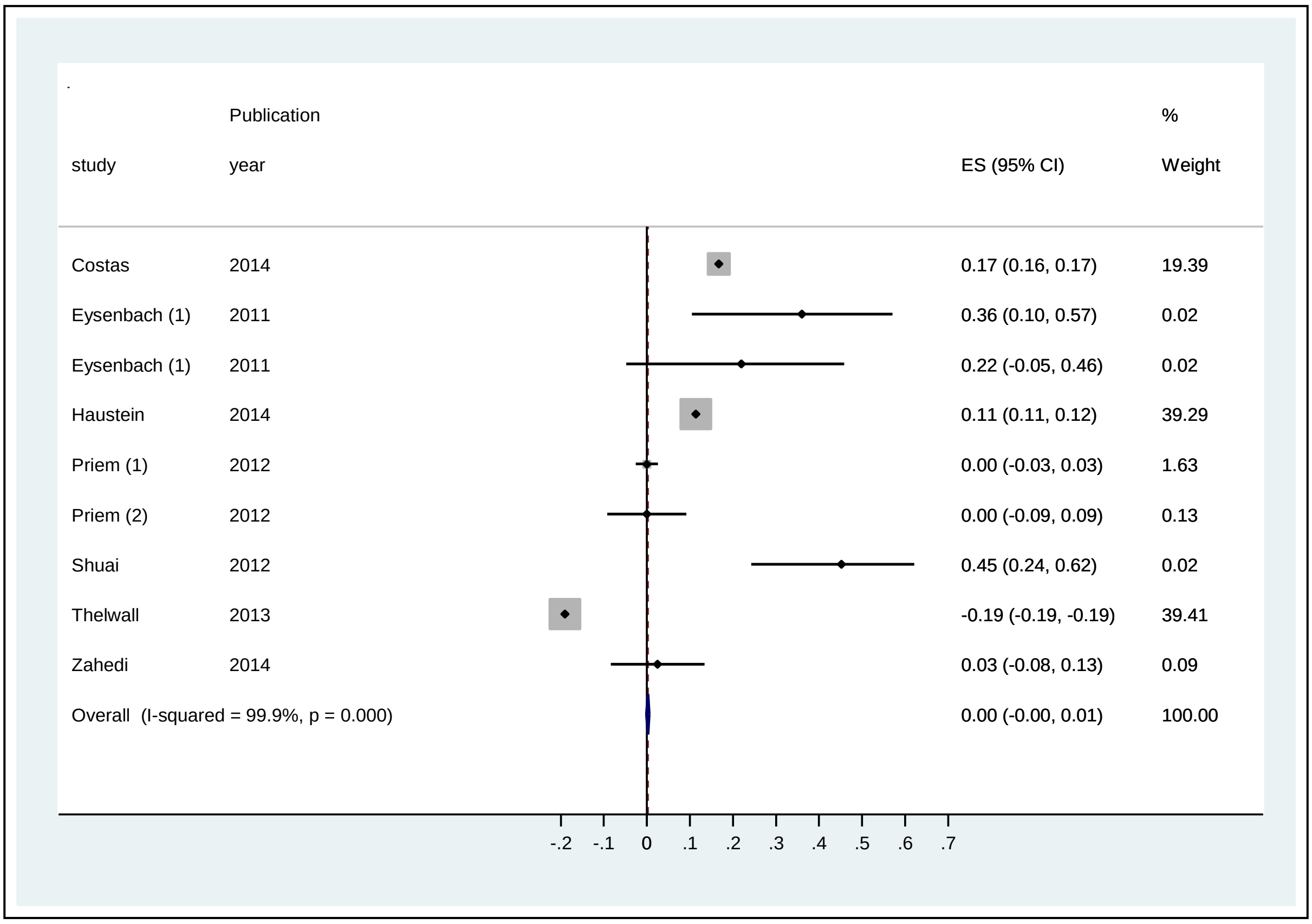

Figure 1. Meta-analysis of correlations between Twitter citations and traditional citations (pooled r=0.003). The correlation coefficients, weighted by sample size of a study/analysis, were included in the meta-analysis.

Most studies into measuring impact by use of Twitter citations have calculated a correlation between traditional citations and Twitter citations. The correlation coefficients from these studies are included in a meta-analysis, the results of which are shown in Figure 1. The analysis included a total of nine coefficients from seven studies, with the following details: Since Eysenbach (2011) calculated a correlation of Twitter counts both with citations from Google Scholar, as well as from Scopus and Priem, Piwowar, and Hemminger (2012) calculated correlation coefficients for two journals, both studies appear twice. Whereas Shuai, et al. (2012) calculated Pearson correlation coefficients on the basis of logarithmic counts, the other studies calculated Spearman correlation coefficients based on the raw counts. Priem, Piwowar, et al. (2012) did not specify which coefficients they calculated. Costas, Zahedi, and Wouters (2014), Haustein, Peters, Sugimoto, et al. (2014), Priem, Piwowar, et al. (2012), Thelwall, Haustein, Lariviere, and Sugimoto (2013) and Zahedi, Costas, and Wouters (2014) used citations from the Web of Science; Eysenbach (2011) and Shuai, et al. (2012) worked with citations from Google Scholar or Scopus.

As the results in Figure 1 show, the studies produced very different coefficients. Higher coefficients result particularly from studies with low case numbers. The pooled coefficient of the meta-analysis is r=0.003: This indicates that in all the studies, Twitter counts and traditional citation counts are not correlated with one another.

### 3.2 Online reference manager (Mendeley and CiteULike)

Online reference managers combine social bookmarking service and reference management functionalities in one platform. They can be seen as the scientific variant of social bookmarking platforms, in which users can save and tag web resources (e.g. blogs or web sites). The best known online reference managers are Mendeley (www.mendeley.com) and CiteULike (www.citeulike.org), which were launched in 2004 (CiteULike) and 2008 (Mendeley), and can be used free of cost (Li, Thelwall, & Giustini, 2012). Mendeley – in 2013 acquired by Elsevier (Rodgers & Barbrow, 2013) – has developed since then into the most popular product among the reference managers (Haustein, 2014), and most empirical studies involving reference managers have used data from Mendeley.

The platforms allow users to save or organise literature, to share literature with other users, as well as to save keywords and comments on a publication or to assign tags to them (Bar-Ilan, et al., 2014; Haustein, Peters, Bar-Ilan, et al., 2014). Tags can be freely selected by the user or adopted from other users; there is also the possibility of crowd-sourced annotations of resources or of social tagging (Haustein, 2014). Even if it is literature that is mainly saved by the users, they can also add to a library other products of scientific work, such as data sets, software and presentations. The providers of online reference managers make available a range of data for the use of publication by the users: The most important numbers are the user counts, which provide the number of readers of publications via the saves of publications (Li, et al., 2012). The readers can be differentiated into status groups. The readers' data from Mendeley is also evaluated to make suggestions to the users for new papers and potential collaborators (Galloway, et al., 2013; Priem & Hemminger, 2010).

Mendeley and CiteULike are used chiefly by science, technology, engineering and mathematics researchers (Neylon, et al., 2014). According to a questionnaire in the bibliometric community (Haustein, Peters, Bar-Ilan, et al., 2014), 77% of those questioned know Mendeley and 73% CiteULike. But Mendeley is actually used by only 26% of those questioned and CiteULike by only 13%. Seen overall, one can assume about 6 million saved papers for CiteULike and about 34 million for Mendeley (figures for the year 2012). However, with respect to the number of saved papers there are large differences between disciplines: For example, only about a third of the humanities articles indexed in the Web of Science can also be found in Mendeley; however, in the social sciences, it is more than half (Mohammadi & Thelwall, 2013). Among the reference managers, Mendeley seems to have the best coverage of globally published literature (Haustein, Peters, Bar-Ilan, et al., 2014; Zahedi, et al., 2014). According to an investigation by Li, et al. (2012), only about 60% of the *Nature* or *Science* papers are stored in CiteULike, but over 90% in Mendeley. The large user population and coverage result in Mendeley being seen as the most promising new source for evaluation purposes among the online reference managers (Haustein, 2014). Priem (2014) sees Mendeley already as a rival to commercial databases, such as Scopus and Web of Science.

With a view to the use of the data from online reference managers in research evaluation, bookmarks to publications (i.e. the saving of bibliographic data about publications in libraries) express the interest of a user in a publication (Weller & Peters, 2012). But this interest is very variable; the spectrum extends from simple saving of the bibliographic data of a publication up to painstaking reading, annotation and use of a publication (Shema, et al., 2014; Thelwall & Maflahi, in press). According to Taylor (2013), the following motives could play a role in the saving of a publication: “Other people might be interested in this paper … I want other people to think I have read this paper … It is my paper, and I maintain my own library … It is my paper, and I want people to read it … It is my paper, and I want people to see that I wrote it” (p. 20). The problem of the unclear meaning of the saving (or naming) of a publication is common to bookmarks in reference managers and also many other traditional and alternative metrics: Traditional citations can mean either simple naming citations in the introduction to a paper, as well as extensive discussions in the results or discussion sections (Bornmann & Daniel, 2008). Traditional citations can also be self-citations.

The data from online reference managers is seen as one of the most attractive sources for the use of altmetrics in research evaluation (Sud & Thelwall, in press). The following reasons are chiefly given for this:

- The collection of literature in reference managers is – similar to the way this is the case with citations and downloads of publications – a by-product of existing workflows (Haustein, 2014). This is why saves are appropriate as an alternative metric chiefly for the measurement of impact in areas of work where literature is collected and evaluated, such as with researchers in academic and industrial research, students and journalists. Saves should therefore be usable as a data source for the measurement of impact in and beyond (academic) research. If impact measurement does not only count the saves, but also takes into account further context via metadata capture and user profiles, a named publication can even be assigned a particular meaning which users connect with it (Gunn, 2013; Haustein, 2014).
- According to Mohammadi and Thelwall (2014), usage data of literature may be partially available (i.e. from publishers); but there is a shortage of global and publisher-independent usage data. This gap could be closed with data from online reference managers whose coverage of literature is already good (Gunn, 2013).

- The frequency of use of a paper can only be adequately assessed as high or low if this number is viewed in relation to a reference set. This is why indicators where citation impact is normalised with the help of a reference set have established themselves in bibliometrics as the normalised indicators for the measurement of citation impact. Working with reference sets is also suggested with regard to usage statistics. Here the number of saves of a paper is seen relative to the number of saves for other papers from the same journal (Haustein, 2014). For the evaluation of journals it has already been suggested to calculate usage ratios for journals, with which the performances of journals can be compared with one another, standardised for size: "Usage Ratio of a journal is defined as the number of bookmarked articles divided by the total number of articles published in the same journal during a certain period of time" (Haustein & Siebenlist, 2011, pp. 451-452).

However, the use of data from online reference managers is not only seen as advantageous, but also as problematic:

- A great advantage assumed for altmetrics compared with traditional citations is that they can be evaluated sooner after the appearance of a publication than citations. In contrast to Twitter citations, which reach the maximum number per day very soon after publication, with saves it takes – according to Lin and Fenner (2013) – several months to accumulate. Thus, even when literature can be saved as soon as it appears (Li, et al., 2012), the data from online reference managers only have an uncertain advantage for use in research evaluation as compared with traditional metrics.

- Whereas there is a dominant platform, namely Twitter, in micro-blogging, with online reference managers there are several (besides Mendeley and CiteULike, also Zotero and RefWorks). Even if Mendeley seems to be developing into a dominant platform, the use of data from Mendeley alone could lead to the problem that literature may be saved on another platform (e.g. CiteULike), but not evaluated for the altmetric in use (Neylon, et al., 2014). Since not everybody who reads and uses scientific literature works with an online reference manager, there is the additional problem that the evaluation of save data only takes into account a part of the actual readership. Among researchers this part is probably younger, more sociable and more technologically-oriented than average for researchers (Sud & Thelwall, in press).
- The data which are entered by users into the online reference managers are erroneous or incomplete. This can lead to saves not being able to be associated unambiguously with a publication (Haustein, 2014).

Many of the empirical-statistical studies into social bookmarking – according to Priem and Hemminger (2010) – deal with tags and tagging. Seen overall, the studies come to the conclusion that “exact overlaps of tags and professionally created metadata are rare; most matches are found when comparing tags and title terms” (Haustein & Peters, 2012). A large part of the studies into online reference managers has evaluated the correlation between traditional citations (from Scopus, Google Scholar and the Web of Science) and bookmarks in Mendeley and/or CiteULike. An overview of the differences between the studies can be found in Table 1. Similar to Twitter citations in section 3.1; in this section too the results of these studies are used to calculate a meta-analysis. However, the meta-analysis omits the few studies which relate to other reference managers, such as Connotea or BibSonomy, or other traditional citations, such as citations recorded by PubMed Central or CrossRef.

Table 1. Studies which were included in the meta-analysis of the correlation between traditional citations and bookmarks in online reference managers

| Abbreviation | Study | Correlation | Platform | Citation source | Sample |
|---|---|---|---|---|---|
| Bar-Ilan (1) | Bar-Ilan (2012a, 2012b) | Spearman | Mendeley | Web of Science | JASIST |
| Bar-Ilan (2) | Bar-Ilan (2012a, 2012b) | Spearman | Mendeley | Scopus | JASIST |
| Bar-Ilan (3) | Bar-Ilan (2012a, 2012b) | Spearman | Mendeley | Google Scholar | JASIST |
| Bar-Ilan1 (1) | Bar-Ilan et al. (2012) | Spearman | Mendeley | Scopus | papers from scientometricians |
| Bar-Ilan1 (2) | Bar-Ilan, et al. (2012) | Spearman | CiteULike | Scopus | papers from scientometricians |
| Haustein (1) | Haustein, Peters, Bar-Ilan, et al. (2014) | unknown | Mendeley | Scopus | papers from scientometricians |
| Haustein (2) | Haustein, Peters, Bar-Ilan, et al. (2014) | unknown | CiteULike | Scopus | papers from scientometricians |
| Henning | Henning (2010) | unknown | Mendeley | Web of Science | top 10 journal articles |
| Li (1) | Li, et al. (2012) | Spearman | Mendeley | Web of Science | *Nature* |
| Li (2) | Li, et al. (2012) | Spearman | Mendeley | Google Scholar | *Nature* |
| Li (3) | Li, et al. (2012) | Spearman | CiteULike | Web of Science | *Nature* |
| Li (4) | Li, et al. (2012) | Spearman | CiteULike | Google Scholar | *Nature* |
| Li (5) | Li, et al. (2012) | Spearman | Mendeley | Web of Science | *Science* |
| Li (6) | Li, et al. (2012) | Spearman | Mendeley | Google Scholar | *Science* |
| Li (7) | Li, et al. (2012) | Spearman | CiteULike | Web of Science | *Science* |
| Li (8) | Li, et al. (2012) | Spearman | CiteULike | Google Scholar | *Science* |
| Li1 (1) | Li and Thelwall (2012) | Spearman | Mendeley | Web of Science | non-anomalous Genomics & Genetics articles |
| Li1 (2) | Li and Thelwall (2012) | Spearman | Mendeley | Google Scholar | non-anomalous Genomics & Genetics articles |
| Li1 (3) | Li and Thelwall (2012) | Spearman | Mendeley | Scopus | non-anomalous Genomics & Genetics articles |
| Li1 (4) | Li and Thelwall (2012) | Spearman | CiteULike | Web of Science | non-anomalous Genomics & Genetics articles |
| Li1 (5) | Li and Thelwall (2012) | Spearman | CiteULike | Google Scholar | non-anomalous Genomics & Genetics articles |
| Li1 (6) | Li and Thelwall (2012) | Spearman | CiteULike | Scopus | non-anomalous Genomics & Genetics articles |
| Li1 (7) | Li and Thelwall (2012) | Spearman | Mendeley | Web of Science | anomalous Genomics & Genetics articles |
| Li1 (8) | Li and Thelwall (2012) | Spearman | Mendeley | Google Scholar | anomalous Genomics & Genetics articles |
| Li1 (9) | Li and Thelwall (2012) | Spearman | Mendeley | Scopus | anomalous Genomics & Genetics articles |
| Li1 (10) | Li and Thelwall (2012) | Spearman | CiteULike | Web of Science | anomalous Genomics & Genetics articles |
| Li1 (11) | Li and Thelwall (2012) | Spearman | CiteULike | Google Scholar | anomalous Genomics & Genetics articles |
| Li1 (12) | Li and Thelwall (2012) | Spearman | CiteULike | Scopus | anomalous Genomics & Genetics articles |
| Liu | Liu, Xu, Wu, Chen, and | Spearman | CiteULike | Scopus | PLoS papers |
| | Guo (2013) | | | | |
| Mohammadi (1) | Mohammadi and Thelwall (2014) | Spearman | Mendeley | Web of Science | social science articles |
| Mohammadi (2) | Mohammadi and Thelwall (2014) | Spearman | Mendeley | Web of Science | humanities articles |
| Priem (1) | Priem, Piwowar, et al. (2012) | unknown | Mendeley | Web of Science | *PLoS One* |
| Priem (2) | Priem, Piwowar, et al. (2012) | unknown | Mendeley | Web of Science | *PLoS Pathogens* |
| Priem (3) | Priem, Piwowar, et al. (2012) | unknown | Mendeley | Web of Science | *PLoS One* |
| Priem (4) | Priem, Piwowar, et al. (2012) | unknown | Mendeley | Web of Science | *PLoS Pathogens* |
| Schlögl | Schlögl, Gorraiz, Gumpenberger, Jack, and Kraker (2014) | Spearman | Mendeley | Scopus | *Information and Management* |
| Schlögl1 | Schlögl, Gorraiz, Gumpenberger, Jack, and Kraker (2013) | Spearman | Mendeley | Scopus | *Journal of Strategic Information Systems* |
| Sud | Sud and Thelwall (in press) | Spearman | Mendeley | Web of Science | Biochemistry papers |
| Weller (1) | Weller and Peters (2012) | Kendall | Mendeley | Scopus | author group 1 |
| Weller (2) | Weller and Peters (2012) | Kendall | CiteULike | Scopus | author group 1 |
| Weller (3) | Weller and Peters (2012) | Kendall | Mendeley | Scopus | author group 2 |
| Weller (4) | Weller and Peters (2012) | Kendall | CiteULike | Scopus | author group 2 |
| Weller (5) | Weller and Peters (2012) | Kendall | Mendeley | Scopus | author group 3 |
| Weller (6) | Weller and Peters (2012) | Kendall | CiteULike | Scopus | author group 3 |
| Yan | Yan and Gerstein (2011) | Spearman | CiteULike | Scopus | PLoS papers |
| Zahedi | Zahedi, et al. (2014) | Spearman | Mendeley | Web of Science | random publications |

As Table 1 shows, a large number of studies have already been published concerning the correlation between traditional citations and bookmarks in online reference managers. The studies differ chiefly in the use of the correlation coefficient (Spearman or Pearson), the reference manager platform, the source of the traditional citations and the sample investigated. Since around half of the studies used each of CiteULike (41%) or Mendeley (59%), the results of the meta-analysis will be presented separately for each platform. The results of the meta-analyses are displayed in Figure 2. Whereas the studies which evaluated data from Mendeley resulted in a pooled r=0.51, for the data from CiteULike the value is r=0.23 (the overall pooled r=0.37). Thus for Mendeley, there is a large effect for the strength of the relationship and for CiteULike a small to medium effect. The higher correlation for the Mendeley data most probably reflects the better coverage of the literature in Mendeley than in CiteULike.

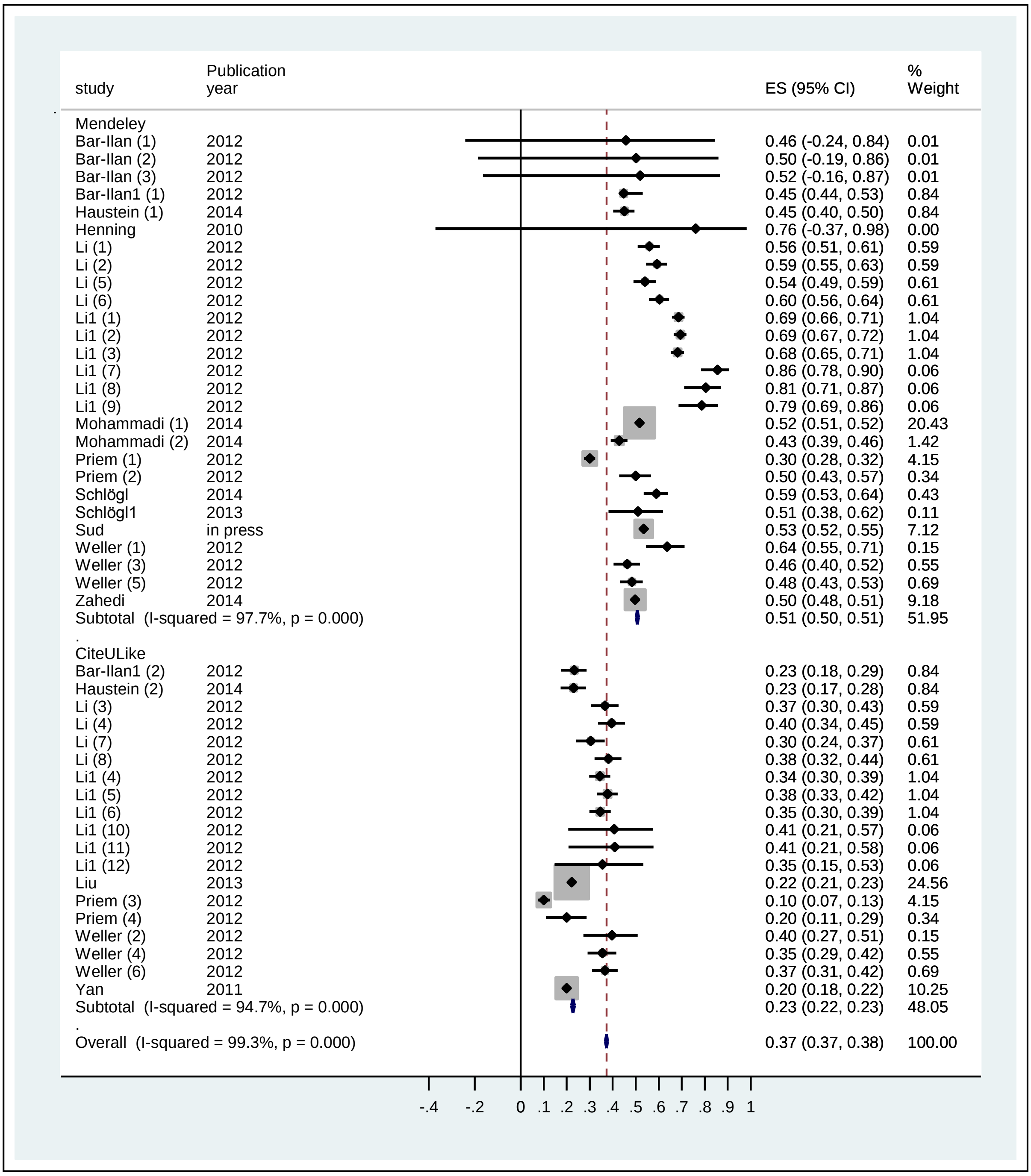

Figure 2. Meta-analysis of correlations between bookmarks in online reference managers (Mendeley and CiteULike) and traditional citations (pooled r=0.37). The correlation coefficients, weighted by sample size of a study/analysis, were included in the meta-analysis.

### 3.3 Blogging

Blogs or weblogs are online forums which are run by one or more people, and (continually) publish contributions on particular topics. They serve as easy-to-publish media and can combine text, images, audio and video content, and links to other blog posts, web pages, and other media (Kovic, Lulic, & Brumini, 2008; Weller & Peters, 2012). In contrast to microblogging (see section 3.1), posts in blogs are not limited to a few characters, but may consist of extensive texts. Blog posts are usually published as informal essays (Mewburn & Thomson, 2013).

Puschmann and Mahrt (2012) define blog posts as scholarly if they are "written by academic experts that are dedicated in large part to scientific content" (p. 171). Scholarly media outlets, e.g. National Geographic, the Nature Group, Scientific American, and the PLoS journals, have science blogging networks (Bar-Ilan, et al., 2014). Increasingly, universities and research institutions offer platforms to publish content from students and faculty members (Puschmann, 2014). Blogs generally offer the possibility of commenting on the posts of bloggers, which may result in extended informal discussions about academic research (Shema, Bar-Ilan, & Thelwall, 2012b).

Whereas most blogs are only read by a few people, there are a few blogs whose posts continue to excite a very lively interest. Since with blogging – unlike with microblogging on Twitter – there is (not yet) a platform that has established itself as dominant, blog posts are grouped according to topic by blog aggregators (Priem, 2014). For example, posts on scientific topics can be found on ResearchBlogging (Shema, Bar-Ilan, & Thelwall, 2012a): "ResearchBlogging.org allows readers to easily find blog posts about serious peer-reviewed research, instead of just news reports and press releases" (researchblogging.org/static/index/page/about). For a post to appear on ResearchBlogging, it must fulfil certain minimum standards, which are checked by a moderator, such as, for example, peer-reviewed publications are discussed and at least a full and complete citation is included (Fausto et al., 2012; Groth & Gurney, 2010). Other blog aggregators besides ResearchBlogging are Nature Blogs and ScienceSeeker.

Bloggers who write about scientific topics mainly come from the academic environment; many have a doctorate (Shema, et al., 2014). According to Bonetta (2007) (citing a blog of Bora Zivkovic), science blog posts are mainly "written by graduate students, postdocs and young faculty, a few by undergraduates and tenured faculty, several by science teachers, and just a few by professional journalists" (p. 443). The boundaries between science journalism and science blogging are blurred (Kovic, et al., 2008; Puschmann, 2014). Generally, those academics blog who are also journalistically active. Here it could be a case of journalists who have specialised in scientific topics, or also of researchers who also write journalistically. Besides individuals, scientific organisations and science-related journals now also blog (Puschmann & Mahrt, 2012).

As a rule, science blogs either discuss particular new results published in academic journals, or scientific themes are discussed which are of special interest to the public (e.g. global warming) (Groth & Gurney, 2010). However, the blogs do not just deal with research results themselves, but also many topics associated with science, such as the relation between science and society, a researcher's life, and problems of academic life (Colson, 2011; Wolinsky, 2011). Blog posts about scientific topics are chiefly read by readers with an academic background and science journalists, but also by laypersons (Mewburn & Thomson, 2013; Puschmann & Mahrt, 2012).

If blogs are used as a source for altmetrics, it is usually a matter of the naming of papers in the blog posts (Bar-Ilan, et al., 2014). According to Lin and Fenner (2013) for example, 5% of the papers publicised by PLoS are discussed in science blogs. With regard to the naming of papers in blog posts, Shema, et al. (2014) distinguish between blog mentions and blog citations: Whereas blog citations mean the citation of scholarly material in structured, formal styles, blog mentions mean any kind of naming of scholarly material.

Which advantages do blogs offer that make them appear attractive as sources for altmetrics?

- The most important advantage of blogs is seen as their social function, to offer and explain scientific material to the general public (Fausto, et al., 2012). With its wide-ranging dialogue, science blogging can build a bridge between research and other parts of society (Colson, 2011; Kjellberg, 2010). If one assumes that in the USA around a third of the population read blog posts (Puschmann & Mahrt, 2012), one can really reach a wide audience with blogging. Blogs could then play an important role in the generation of societal impact by research (Fausto, et al., 2012). Whereas the results from Mewburn and Thomson (2013) show that the transfer of research into society was one of the less popular motivations for academics to blog, the results from Colson (2011) still suggest that science bloggers want to communicate their scientific knowledge to the general public.
- Unlike microblogs, blogs allow one to treat particular contents extensively. A good example of extensive treatment of a topic are the posts concerned with the reasons for, and the effects of, global warming (Thorsen, 2013). The extensive treatment of topics facilitates knowledge transfer from science into society, since the research results are appropriately presented for people outside science. Only when the research results are understood and their social relevance is recognised can the transfer into society take place.

  Many science bloggers therefore also write for a lay public, since they are no longer satisfied with the quality of reporting in traditional journalism – chiefly due to the lack of understanding, the over-simplification and the sensationalism (Colson, 2011). Well written and high quality blog posts are, however, used by the traditional media as input for their own reports (Bonetta, 2007).

– Blog posts are seen as a new possibility for post-publication peer review, where the quality of publications (their importance, correctness and relevance) with regard to their use in particular segments of society can be tested (Fausto, et al., 2012). Blogs could allow a quick forum for public peer review of research to be set up (Thorsen, 2013). This would not be a matter of abolishing traditional peer review, but of complementing the present system (Shema, et al., 2014). The possibility is seen as particularly interesting in this connection that people who would not normally be regarded as "peers" could review a paper (Shema, et al., 2012a).

However, the use of blogs as sources for altmetrics does not only have advantages, but also disadvantages, the most important of which are listed as follows:

– Whereas many Web 2.0 social media are associated with one or a few platforms (as with micro-blogging and Twitter), blogs are distributed over the whole web (Priem & Hemminger, 2010). That significantly complicates the collection of blog citations. Aggregators are a help in this connection, but not a completely satisfactory solution, since they scarcely include more than a subset of all the blogs available (Shema, 2014).
– Unlike citations in journals, which are recorded relatively reliably in the multidisciplinary databases (e.g. Web of Science), blog citations are transient and links obsolesce with time (Shema, et al., 2012a). "For example, blogs move to a blog network or leave it, become invitation-only, or disappear from the Web altogether" (Shema, et al., 2012b, p. 190). In this connection, Bar-Ilan, et al. (2014) report about a list of 50 popular science blogs which were publicised in 2006 in *Nature*. Six years later, two of these blogs could no longer be accessed, 16 were inactive and three were hibernating. Overall, only about half of the blogs could still be accessed.

- For blog posts there are no guidelines as to how publications should be cited. This results in very disparate forms for literature citations, which makes it very difficult to recognise citations of a particular publication. Aggregators, such as ResearchBlogging.org, try to counter this problem by the enforcement of standards in posts (Shema, et al., 2014).
- According to Wolinsky (2011), science blogging is still in the "hobby zone." Most bloggers earn nothing at all, or only very little, through blogging. Against this background one cannot expect the general quality of blog posts to be very high (Bar-Ilan, et al., 2014). Editors are therefore regarded as important for the quality control of posts (Thorsen, 2013).

Besides the studies on blogs which address the relation between blog citations and traditional citations (and which are discussed further below), a series of studies which focus on other topics could be investigated. A selection of these studies is presented in the following.

Groth and Gurney (2010), for example, have evaluated 295 chemistry blog posts written about peer-reviewed literature. As the results of the study show, blog posts are more immediate and contextually relevant than traditional academic literature. They are more concerned with the non-technical implications of science, and focus on publications which have appeared in high-impact journals. The focus of the bloggers on publications from high-impact journals could also be demonstrated in another study: Shema, et al. (2012a) investigated a sample of blog posts from ResearchBlogging.org and observed that the papers referred to, from the areas of life and behavioural sciences, were mainly from *Science*, *Nature*, *Proceedings of the National Academy of Sciences* (PNAS), and *PLoS One*.

The underlying data for the study of Luzón (2013) was constituted by 75 blog posts from 15 different blogs. The posts were investigated with a view to their rhetorical structure. The analysis revealed "that science bloggers deploy a variety of rhetorical strategies to contextualise scientific knowledge for a diverse audience and situate this knowledge in public life, thus helping the public make informed decisions, but also to position themselves regarding the research reported and to persuade the readers of their own interpretation, especially in controversial issues regarding civic life" (Luzón, 2013). Shema, Bar-Ilan, and Thelwall (in press) have investigated 391 blog posts from 2010 to 2012 in Researchblogging.org's health category, with the help of content analysis methods. Their results show that "health research bloggers rarely self-cite and the vast majority of their blog posts (90%) include a general discussion of the issue covered in the article, with over a quarter providing health-related advice based on the article(s) covered. These factors suggest a genuine attempt to engage with a wider non-academic audience. Nevertheless, almost 30% of the posts included some criticism of the issues being discussed".

Let us now come to the results of the meta-analysis on the correlations between blog counts and traditional citation counts. Figure 3 represents the results of the meta-analysis. The analysis covered the data from five studies: Costas, et al. (2014), Liu, et al. (2013), Priem, Piwowar, et al. (2012), Thelwall, et al. (2013), and Yan and Gerstein (2011). Three of these studies are based on article-level metrics, which are offered by PLoS (Liu, et al., 2013; Priem, Piwowar, et al., 2012; Yan & Gerstein, 2011). All three studies use Scopus as the source for the traditional citations. The two other studies (Costas, et al., 2014; Priem, Piwowar, et al., 2012), included in the meta-analysis, use data from the Web of Science. The meta-analysis omits the few analyses based on citations from other sources (such as PubMed Central or CrossRef). In the framework of their article-level metrics, PLoS offers blog counts from Nature Blogs, Bloglines and ResearchBlogging.org (Liu, et al., 2013; Priem, Piwowar, et al., 2012; Yan & Gerstein, 2011). As the results of the meta-analysis in Figure 3 show, for all studies the overall pooled r=0.12. This indicates a low correlation between blog counts and traditional citations.

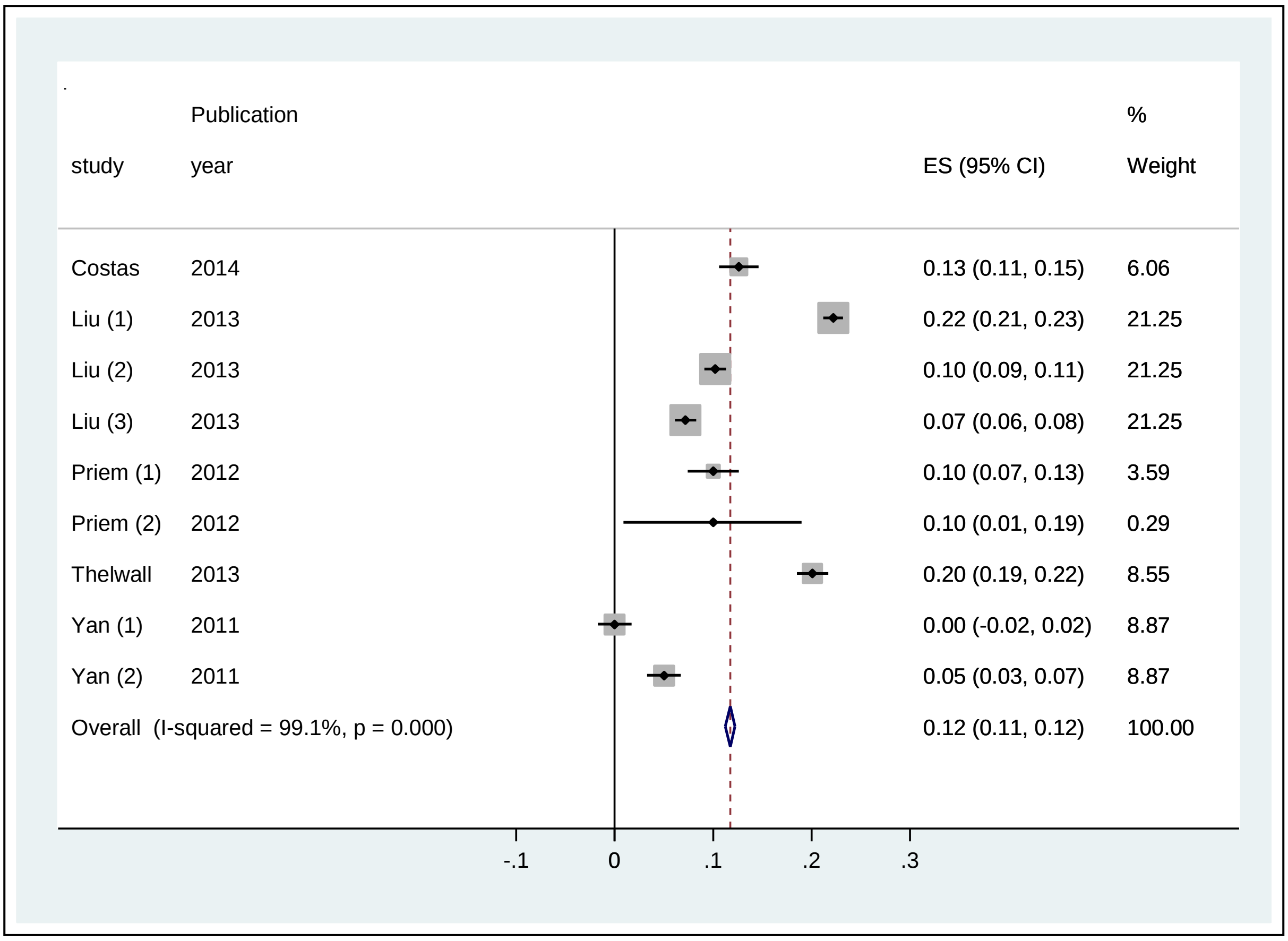

Figure 3. Meta-analysis of correlations between blog counts and traditional citation counts (pooled r=0.12). The correlation coefficients, weighted by sample size of a study/analysis, were included in the meta-analysis.

# 4 Discussion

Alternative metrics are currently one of the most popular research topics in scientometric research. This topic seems on the way to replacing h index research, which dominated in recent years, as the most popular research topic. Both at the relevant conferences, such as the ISSI conference, and in the most important journals, such as *Scientometrics*, *Journal of Informetrics* and the *Journal of the American Society of Information Science and Technology*, more and more contributions on this topic are being published. There are two important reasons for the popularity of this topic: On the one hand, the data available on the social media platforms provides an easily accessible source for statistical analyses, which is eagerly adopted by the research community. Meaningful impact data for a large publication set are normally not easy to obtain in bibliometrics. On the other hand, tapping this new data source matches the desire in science policy to measure the broad impact of research. Currently it is under investigation in scientometrics whether this desire can be fulfilled with altmetrics.

The earlier sections of this study provided an overview of research on microblogging (Twitter), online reference managers (Mendeley and CiteULike) and blogging. This overview was compiled chiefly before the background of the question as to how applicable these altmetrics are for the evaluation of research. Thus, for each altmetric the related advantages and disadvantages are discussed which should be considered for their use in research evaluation. Since research in recent years has concentrated on the correlation of altmetric counts with citation counts, the overview also focused on this topic. For each altmetric, a meta-analysis was calculated for the related correlation coefficients, in order to arrive at a general opinion. As the results of the meta-analyses show, the correlation with traditional citations for micro-blogging counts is negligible (pooled r=0.003), for blog counts it is small (pooled r=0.12) and for bookmark counts from online reference managers, medium to large (CiteULike pooled r=0.23; Mendeley pooled r=0.51).

Twitter and blog citations seem to measure something different from traditional citations, because the pooled coefficients are very low. A similar result – with regard to Twitter – was already reached by Bornmann (2014). An alternative explanation for the absence of a correlation or a low correlation is that both altmetrics have little value. For instance, if tweets are generated completely at random then they would have a correlation of close to zero with citation counts. The greater the element of randomness in tweets, the lower the expected correlation with citations if other factors are equal.

The results of the meta-analyses have several limitations which should be considered in their interpretation:

1) Even if the results of the meta-analysis indicate that Twitter and blog citations measure something different compared with traditional citations, the meta-analysis still does not allow us to answer the question of an added value. It is not clear which impact of research is measured when Twitter citations are collected for a publication set. It does seem to be a different construct than measured by citations, but it is not clear what exactly this other construct is (if any, see above). Thus, even if the meta-analysis leads to interesting results, from these results new research questions arise which should be clarified before altmetrics are used in the evaluation of research. Such an indication of research questions which need clarification can be found in almost all publications which deal (empirically) with altmetrics. One of the most important publications in this connection came from NISO Alternative Assessment Metrics Project (2014). This project involved compiling a white paper on altmetrics, in which a total of 25 action items in 9 categories were identified. Two examples of these action items are: “Develop specific definitions for alternative assessment metrics” and “Identify specific scenarios for the use of altmetrics in research evaluation (e.g., research data, social impact) and what gaps exist in data collection around these scenarios.”

2) Meta-analyses are appropriate when the data come from situations where the results can be broadly expected to be the same and so combining different studies would give the best estimate of the “true effect size”. This is not the case in scientometrics (altmetrics and bibliometrics). Substantial differences between fields and years in the relationship between citations and many other variables (e.g., co-authorship and nationality of authors) can be expected and so it is problematic to average results between fields or years. This issue of changes over time is also particularly strong for any social web variable, because the uptake of the social web has increased rapidly over the past half-decade. This might produce major qualitative differences in the way in which people use different sites. In future meta-analytic studies, field- and time-dependencies of the data should be considered.

3) Knowing the pooled correlation coefficient for a group of studies is typically the starting point of a further meta-analysis (Marsh, Bornmann, Mutz, Daniel, & O'Mara, 2009). Since there is systematic variation among the pooled studies beyond what can be explained by sampling variability (which is almost always the case), it is important to determine in a next step what study characteristics (sample, design etc.) are able to explain study-to-study variation in the results.

4) There are several dependencies in the underlying datasets which have not been considered in the meta-analyses of this study. First, bibliometric analyses are typically based on the same source. Indeed, while meta-analyses in the medical sciences pool studies that have typically been made using different samples (patients), in bibliometrics, the same source (e.g. papers indexed in Web of Science) is typically used in several pooled studies, resulting in the double (triple, quadruple, etc.) counting of some cases and, thus, increase the weight of these cases in the pooled sample. Second, (i) the different pooled studies in this meta-analysis used the same data set, (ii) the same dataset is analysed several times in one and the same study, or (iii) a study used different datasets to produce several correlation coefficients. For example, Li and Thelwall (2012) analysed the same dataset twice, once with and once without outliers. Priem, Piwowar, et al. (2012) used several datasets from PLoS and reported several correlation coefficients. Sometimes, these dependencies in the underlying datasets of the pooled studies are difficult to detect. Since the weight of cases that are counted twice or more is overestimated in a meta-analysis (Tramer, Reynolds, Moore, & McQuay, 1997), the different dependencies between the studies should be tried to uncover und should be considered in future meta-analytic models of altmetrics data (Cheung, 2014).

5) The goal of meta-analyses is to highlight relationships that would otherwise be invisible because of size effects. In other words, by increasing the size of the sample, we are able to observe “new” results and, thus, the combined results become greater than the sum of its parts. In the meta-analyses of this study, there are large-scale analyses that have been done already, and all of them point in the same direction (no correlation for Twitter counts, small correlation for blog counts, and medium to large correlation for bookmark counts). It is not as if all existing studies were based on small (independent) publication sets, in which case a meta-analysis might be useful. But in the current case, existing studies already provide strong evidence about the altmetrics versus citation relationship, and the meta-analysis can add only little to the current body of knowledge. In contrast to medical studies, bibliometric and altmetric analyses are often carried out on a certain population of papers (i.e. all peer-reviewed journal articles in biomedicine) instead of small samples.

# 5 Conclusions and future research

The meta-analyses of this study have shown (as expected) that the more a social media community is dominated by people focussing on research, the higher the correlation between the corresponding altmetric and traditional citations is. For example, since the use of reference managers for the organization of research publications can be seen as a typical feature of people working research-affine, bookmark counts have shown the highest correlation with traditional citations. With micro-blogging, the situation is reversed: a low pooled correlation seems to reflect an impact of research on people working research-non-affine.

The results of the frequent studies measuring correlations between altmetrics and traditional citations (and the results of this meta-analysis) should be seen as a first introducing step in research on altmetrics. They provide a good opportunity to focus the field (and the extensive amount of research going into altmetrics) into more profitable avenues. Low correlations point to altmetrics which might be of special interest for the broad impact measurement of research, i.e. impact on other areas of society than science. Future studies should focus on these altmetrics to investigate their specific potential to measure broad impact: Who are the users of research papers outside academia? Where do they work? How do they use research papers? Do they prefer certain scientific literature (e.g. rather reviews than articles)? Are there already altmetrics data available, which allow an impact measurement on a precisely defined user group (e.g. politicians in a certain country)?

# Acknowledgements

I would like to thank Hadas Shema for discussing the concept of this paper.